\documentclass[aps,floatfix,prb,superscriptaddress,twocolumn,longbibliography]{revtex4-2}
\usepackage[english]{babel}
\usepackage{graphicx}
\usepackage{stmaryrd}
\usepackage{amssymb}
\usepackage{amsfonts}
\usepackage{amsmath}
\usepackage{dsfont}
\usepackage{xcolor}
\usepackage{xspace}
\usepackage{multirow}
\usepackage{comment}
\usepackage{url}
\usepackage{hyperref}
\usepackage{bm}
\usepackage{amsthm}

\DeclareSymbolFont{largesymbols}{OMX}{cmex}{m}{n}
\newcommand{\va}{\mathbf{a}}
\newcommand{\vk}{\mathbf{k}}

\newcommand{\be}{\begin{eqnarray}}
\newcommand{\ee}{\end{eqnarray}}

\begin{document}

\title{Flat-band Ferromagnetism of SU$(N)$ Hubbard Model on the Kagome Lattices}

\author{Hao Jin}
\affiliation{College of Physics, Sichuan University, Chengdu, Sichuan 610064, China}

\author{Wenxing Nie}
\email{wxnie@scu.edu.cn}
\affiliation{College of Physics, Sichuan University, Chengdu, Sichuan 610064, China}

\begin{abstract}

The kagome lattice, a well known example of the geometrically frustrated system, hosts a dispersionless flat band that offers a unique platform for studying correlation-driven quantum phenomena. At appropriate particle concentrations, the existence of a flat band allows a representation of percolation with nontrivial weights. In this work, we investigate the paramagnetic-ferromagnetic transition in the repulsive SU($N$) Hubbard model on the kagome lattice within this percolation framework. In this representation, the model can be rigorously mapped to a classical $N$-state site-percolation problem on a triangular lattice, with the SU($N$) symmetry reflected in the nontrivial weights. By large-scale Monte Carlo simulations for SU($3$), SU($4$), and SU($10$) symmetries, we demonstrate that the critical particle concentration for ferromagnetism exceeds the standard percolation threshold and increases with $N$, indicating a strengthening of the effective entropic repulsion.

\end{abstract}

\maketitle

\section{Introduction}
\label{sec:Introduction}

Itinerant ferromagnetism is one of the fundamental topics in condensed matter physics~\cite{Lieb-Mattis,Nagaoka, Lieb-theorem,Mielke1, Mielke2, Mielke3, TasakiPRL, Mielke-Tasaki,  Katsura-Tasaki,MaksymenkoPRL,Katsura-nagaoka-SU(N),YiLi-PRL2014, Wu-QMC-thermo-PRX2015,Nie-PRA, Ruijin, YiLi-PRB2021,Katsura-SU(N)-2021,Katsura-SU(N)}, which is generally regarded as a result of the subtle interplay between Coulomb interaction and many-body effects~\cite{Tasaki-review}. Given that paramagnetism is an inevitable consequence of the Pauli exclusion principle in non-interacting systems, ferromagnetism originates from electron-electron correlations. However, strong interaction is not a sufficient condition, as it has been proven that the absence of ferromagnetism in one-dimensional systems with only nearest-neighbor hoppings, irrespective of the interaction strength~\cite{Lieb-Mattis}. The stabilization of ferromagnetism~\cite{Riera-ED,Brunner-MC,Becca-VMC,HongYao_Nagaoka2D,Wu-QMC-thermo-PRX2015} in itinerant electron systems remains a persistent challenge in condensed matter physics. This difficulty is partly due to the notorious fermion sign problem, which renders rigorous proofs and sign-problem-free numerical simulations of ferromagnetism exceptionally rare and valuable.

The seminal work by Mielke (and independently Tasaki) rigorously proved the existence of unique ferromagnetic ground states in the Hubbard model on a class of lattices featured by a dispersionless (``flat") band, which is referred to as flat-band ferromagnetism~\cite{Mielke1,Mielke2,Mielke3,TasakiPRL,Mielke-Tasaki}. 
Mielke’s lattices are rooted in graph theory~\cite{Mielke1,Mielke2,Mielke3,Mielke-Tasaki}, while Tasaki’s are constructed via geometric cell constructions~\cite{TasakiPRL,Mielke-Tasaki}. As the quenched kinetic energy makes interactions the dominant energy scale, flat band physics can give rise to a wealth of exotic quantum states of matter, including fractional Quantum Hall type states and Wigner crystals~\cite{Wu-WignerCrystal,Neuper-PRL2011,SunPRL,Wen-PRL2011,Bernevig-PRX2011}, with flat-band ferromagnetism being a prominent example. A flat band helps realize ferromagnetism by effectively reducing the interaction energy cost to zero for a spin-polarized state. Such bands typically consist of a large set of degenerate, spatially localized states—a direct consequence of wavefunction localization via destructive quantum interference~\cite{Schulenburg-PRL2002,Zhitomirsky-Tsunetsugu,Derzhko2007summary,DerzhkoPRB2007,Bergman-PRB2008,Altman-PRB2010}.
 
\begin{figure}[hb]
\begin{center}		
\includegraphics[width=0.8\columnwidth]{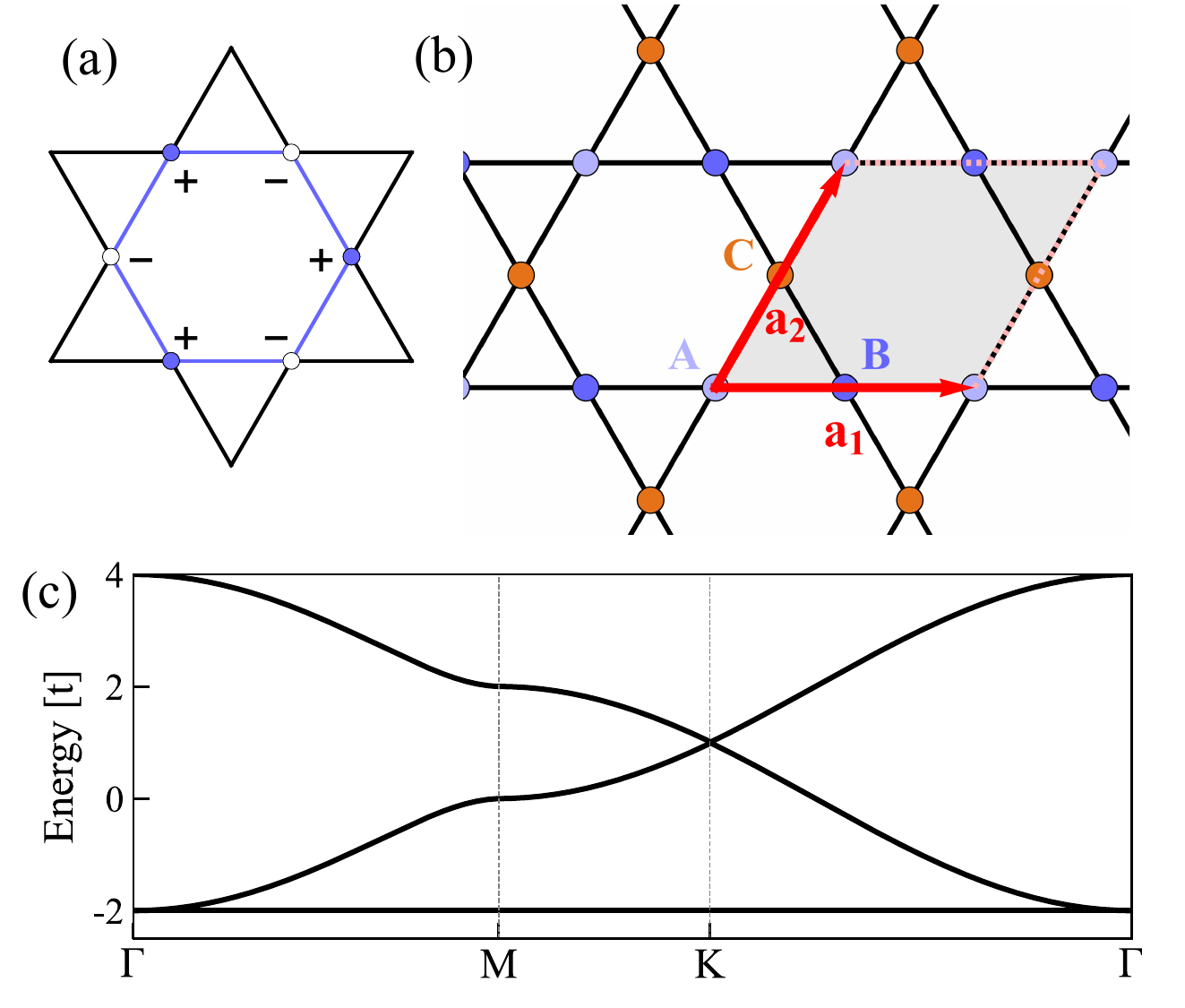}
\end{center}
\caption{(a) Localized hexagon state on the kagome lattice, where the signs $+$ and $-$ label the wavefunction amplitude on each site around the hexagon. 
(b) A sketch of kagome lattice with lattice vectors $\va_1$ and $\va_2$, and the basis sites $A$, $B$, and $C$ in each unit cell (gray). 
(c) Single-particle dispersion of kagome lattice along high-symmetric lines in the Brillouin zone.}
\label{fig:kagome_lattice}
\end{figure}
 
Mielke's proof of flat-band ferromagnetism is based on graph theoretical notion~\cite{Mielke1,Mielke2,Mielke3}. We summarize the main conclusion here. Let $G=(V, E)$ be an abstract graph, where $V$ is the set of vertices and $E$ is the set of edges. One can construct the corresponding line graph $L(G)=(V_L, E_L)$ by taking edges in graph $G$ as vertices in $L(G)$, and identifying an edge in $L(G)$ if the corresponding edges in $G$ have a common vertex. By this definition, the kagome lattice [Fig.~\ref{fig:kagome_lattice} (b)] is the line graph of the hexagonal lattice. 
Mielke proved that for any $U>0$, with $N_e=M(G)$, the ground states of the Hubbard model defined on the line graph $L(G)$ have total spin $S_{\text{tot}}=S_{\max}(=N_e/2)$, and are nondegenerate apart from the  $2S_{\max}+1$-fold degeneracy, where $N_e$ is the number of electrons, $|E|$ ($|V|$) is the number of edges (vertices) of graph $G$, and $M(G)=|E|-|V|+1$ if $G$ is bipartite, and $M(G)=|E|-|V|$  if $G$ is non-bipartite. From Mielke's theorem, we can find that the kagome lattice exhibits ferromagnetism when the filling factor $\nu:=N_e/(2N_{\text{site}})=1/6$ for any $U>0$~\cite{Tasaki-review}, where $N_{\text{site}}$ is the number of lattice sites. For convenience in our percolation analysis, we introduce the particle concentration $p:= N_e/N_{\text{trap}}$, where $N_{\text{trap}}$ is defined as the number of trapping cells in Sec.~\ref{sec:Mapping}. Given that $N_{\text{site}} = 3N_{\text{trap}}$ for the kagome lattice (see Fig.~\ref{fig:mapping_scheme}), we have $\nu = p/6$, so that Mielke's theorem rigorously proved flat-band ferromagnetism appears precisely at particle density $p=1$ for SU($2$) system.
 
The previous rigorous results on itinerant ferromagnetism~\cite{Nagaoka, Mielke1,Mielke2,Mielke3,TasakiPRL,Mielke-Tasaki, Lieb-theorem, YiLi-PRB2021}, including Mielke's theorem~\cite{Mielke1,Mielke2,Mielke3}, 
have been established primarily for traditional electron systems with SU($2$) spin-rotational symmetry. One interesting question is how the ground states of the SU($N$) Hubbard model behave. Generalizing flat-band ferromagnetism to systems with SU($N>2$) symmetry is theoretically intriguing~\cite{Katsura-PRB2019,Cazalilla,Cazalilla_2023,Katsura-PRB2019,Katsura-nagaoka-SU(N),Nie-PRA, Ruijin, Katsura-SU(N)-2021,Katsura-SU(N)} and experimentally relevant~\cite{Takahashi_Nat,Bloch,TaiePRL2010,Hofrichter,Sr-proposal}, owing to the advent of cold atomic gases~\cite{Bloch-RMP}.
In such platforms, atoms with high nuclear spin degeneracy—such as $^{173}$Yb (SU($6$))~\cite{TaiePRL2010,Takahashi_Nat,Taie,Hofrichter} and $^{87}$Sr (SU($10$))~\cite{Sr-Rice,Sr-proposal}—enable the emulation of the SU($N$) fermionic Hubbard model with exceptional controllability and tunability~\cite{Cappellini,ZhangSci,Pagano2014,Scazza}.

The rigorous extension of itinerant ferromagnetism to SU($N>2$) symmetries remains a significant theoretical challenge, leading most existing studies to rely on numerical approaches~\cite{Nie-PRA, Ruijin,Cazalilla_2023,Fuji-PRB-2025,Fuji-arxiv-2025} with few rigorous results~\cite{Katsura-nagaoka-SU(N),Katsura-PRB2019,Katsura-SU(N),Katsura-SU(N)-2021}. For instance, Ref.~\cite{Nie-PRA} provides numerical evidence for a ferromagnetic ground state in the SU($3$) Hubbard model on the Lieb lattice, aiming to generalize Lieb's theorem~\cite{Lieb-theorem}. While Ref.~\cite{Ruijin} examines paramagnetic-ferromagnetic transitions and SU($N$) extension of Tasaki flat-band ferromagnetism~\cite{TasakiPRL} on decorated lattices, using the classical Monte Carlo method. The feasibility of classical Monte Carlo simulations for the SU($N$) Hubbard model hinges on a crucial mapping: from the strongly correlated \emph{quantum} model to a \emph{classical} geometric site-percolation problem~\cite{Mielke-Tasaki,MaksymenkoPRL,YiLi-PRB2021}. This profound connection was first noted by Mielke and Tasaki~\cite{Mielke-Tasaki,Tasaki-review}, who emphasized the necessity of incorporating nontrivial statistical weights, reflecting the Pauli exclusion principle. This idea was initially implemented in Monte Carlo simulations of the SU($2$) system with nontrivial weights~\cite{MaksymenkoPRL,YiLi-PRB2021} and has since been effectively extended to investigate SU($N$) models on Tasaki lattice~\cite{Ruijin}.

In this work, we investigate the paramagnetic-ferromagnetic transition of  SU($N$) Hubbard model on the kagome lattice, aiming to generalize the Mielke flat-band ferromagnetism~\cite{Mielke1,Mielke2,Mielke3}. 
The rest of this paper is organized as follows. In Sec.~\ref{sec:Mapping}, we map the repulsive SU($N$) Hubbard model on the kagome lattice to a classical percolation problem on an effective triangular lattice. In Sec.~\ref{sec:transition}, we investigate the paramagnetic-ferromagnetic transition for systems with SU($3$), SU($4$) and SU($10$) symmetries by large-scale Monte Carlo simulations. The paper is briefly summarized in Sec.~\ref{sec:Conclusion}.

\section{$N$-state Pauli correlated percolation problem}
\label{sec:Mapping}

In this section, we show that the strongly correlated quantum SU($N$) Hubbard model defined on a kagome lattice can be mapped to a classical $N$-state Pauli correlated percolation problem on a triangular lattice, with certain filling restrictions. 

The kagome lattice, composed of corner-sharing triangles [Fig.~\ref{fig:kagome_lattice} (b)], is a paradigmatic platform in condensed matter physics due to its inherent geometric frustration. This specific geometry produces a perfectly flat band in the tight-binding spectrum [Fig.~\ref{fig:kagome_lattice} (c)], which in turn fosters strong electron correlations and topological band structures, making kagome materials a highly promising platform for exploring correlated and topological quantum phenomena~\cite{QSL-review,kagome-Chern,Liu2018,Wang2018}, as evidenced by recent breakthroughs such as the discovery of unconventional superconductivity and competing orders in the AV$_3$Sb$_5$ family (A=K, Rb, Cs)~\cite{Z2-kagome-SC,Mielke-SC,Mielke-SC-nature,Chen-PDW}.

We start with the single-band SU($N$) Hubbard model defined on a kagome lattice as illustrated in Fig.~\ref{fig:kagome_lattice} (b), 
\begin{equation}
H = \sum_{\sigma} \sum_{\langle i,j \rangle} t_{ij} \left( c_{i,\sigma}^{\dag}c_{j,\sigma}+\text{H.c.} \right) + U\sum_{i} \sum_{\sigma \ne \sigma'} n_{i,\sigma}n_{i,\sigma'},
\label{hamiltonian}
\end{equation}
where $c_{i,\sigma}^{\dag}$ ($c_{i,\sigma}$) creates (annihilates) a fermionic particle with flavor (or color) index $\sigma \in \{1, \dots, N\}$ at site $i$, which belongs to a finit lattice $\Lambda$. The operator $c_{i,\sigma}$ obeys the fermionic anticommutation relations $\{ c_{i,\sigma}^{\dag}, c_{j,\sigma'}\}=\delta_{ij}\delta_{\sigma\sigma'}$ and $\{ c_{i,\sigma}, c_{j,\sigma}\}=\{ c_{i,\sigma}^{\dagger}, c_{j,\sigma}^{\dagger}\}=0$. 
For simplicity, the hopping amplitude $t_{ij}$ is restricted between the nearest-neighbor sites, and $t_{ij}=t>0$ is assumed to be uniform. The on-site interaction $U$ is repulsive ($U>0$), and $n_{i,\sigma} = c_{i,\sigma}^{\dag} c_{i,\sigma}$ is the local particle number operator.

The band structure of the corresponding tight-binding model on the kagome lattice consists of a single flat band with energy $\epsilon_0=-2t$ and two dispersive bands with $\epsilon_{\pm} (\vk)=t [1\pm\sqrt{3+2\Gamma(\vk)}]$, as shown in Fig.~\ref{fig:kagome_lattice} (c), where $\Gamma(\vk)=\cos (\vk \cdot \va_1)+\cos(\vk \cdot \va_2)+\cos(\vk \cdot \va_3)$, with lattice vectors $\va_1=a(1,0)$, $\va_2=a(1/2,\sqrt{3}/2)$, $\va_3=\va_1-\va_2$, and $a$ the lattice constant.

The perfectly flat band at the bottom of the single-particle dispersion is the well-known feature of the kagome lattice. The existence of a flat band allows the construction of localized eigenstate of the same energy~\cite{Schulenburg-PRL2002,Zhitomirsky-Tsunetsugu,Derzhko2007summary,DerzhkoPRB2007,Bergman-PRB2008,Altman-PRB2010}, which is confined within a single hexagon as shown in Fig.~\ref{fig:kagome_lattice} (a),
\be
A_{\alpha}^{\dagger}=\frac{1}{\sqrt{6}}\sum_{i^{(\alpha)}=1}^6 (-1)^{i^{(\alpha)}}c_{i^{(\alpha)}}^{\dagger}.
\label{eq:localized}
\ee
Here $\alpha$ is the label of a hexagon, and $i^{(\alpha)}$ enumerates the six successive sites around the $\alpha$-th hexagon. The operator $A_{\alpha}^{\dagger}$ that creates an eigenstate with the energy $-2t$ is localized, as a result of destructive interference on the site adjacent to the hexagon sites.
This can be understood by considering an arbitrary triangle at the boundary of a hexagon. The amplitudes of the three corners of the triangle are $1/\sqrt{6}$, -$1/\sqrt{6}$, and $0$ successively, implying that the hopping amplitude from the first and the second sites onto the third cancel with each other.

Let us first review the mechanism of flat-band ferromagnetism. In the non-interacting limit ($U=0$), the lowest band of the kagome lattice is flat. One can construct real-space localized single-particle eigenstates of this flat band, as shown in Eq.~\eqref{eq:localized}. At low density, the many-particle ground state is obtained by filling the lowest flat band, which results in macroscopic degeneracy, with the ferromagnetic state being one obvious candidate.

\begin{figure}[htpb]
\begin{center}
\includegraphics[width=0.5\columnwidth]{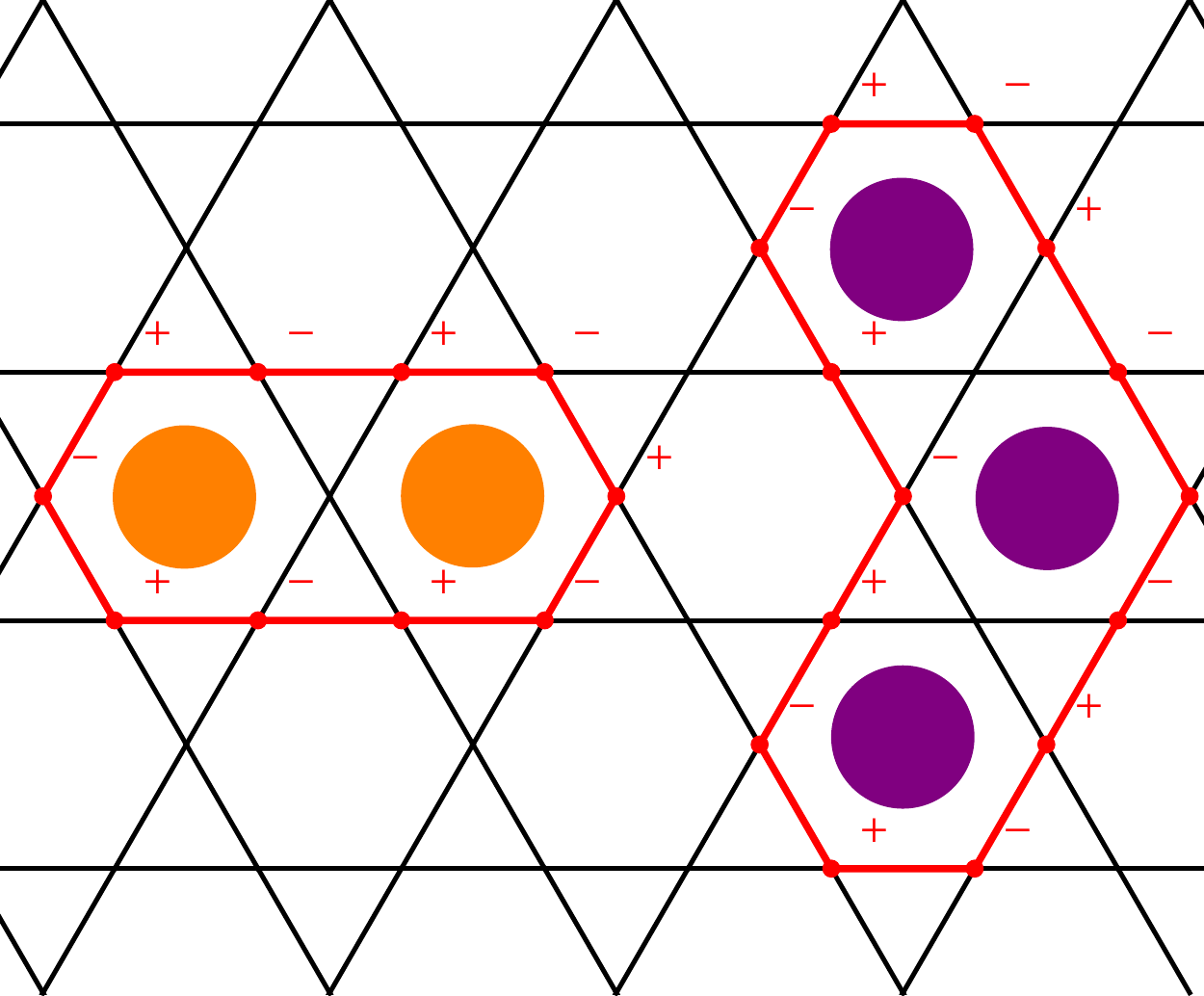}
\end{center}
\caption{An illustration of the mapping from the Hubbard model on a kagome lattice to the percolation problem on an effective triangular lattice. The red lines depict the localized states on the boundary of a double and triple plaquette, with alternating $+$ and $-$ signs labeling the amplitude of the wavefunction on each site. The filled circles depict the linked clusters of the $2$-color percolation model defined on a triangular lattice, which is formed by the centers of the hexagons (trapping cells).}
\label{fig:mapping_scheme} 
\end{figure}

When the repulsive interaction $U>0$ is introduced, states with two or more particles on the same lattice site are penalized by the on-site interaction energy $U$. However, at low particle density [$\nu<1/9$ for SU($2$) case, as discussed in Ref.~\cite{Altman-PRB2010}], electrons can be distributed into isolated hexagons (trapping cells) such that the wavefunctions of adjacent cells do not overlap. As the particle density increases, wavefunctions of electrons in neighboring hexagons inevitably overlap. However, the system can avoid this energy cost by demanding the wavefunction of the adjacent hexagonal trapping cells be fully symmetric in the flavor (spin) part, and by the Pauli exclusion principle, fully antisymmetric in the spatial part. This spatial antisymmetry forces the wavefunction to have nodes at the overlapping sites, thus avoiding the interaction penalty. This mechanism forces all particles within a connected group of occupied trapping cells to align their flavors, forming what are known as ``polarized clusters" (or magnetic clusters). 
The magnetic clusters of increasing size emerge with increasing electron density, leading to a finite magnetic moment. For a restricted range of filling, the interaction $U$ completely lifts the ground-state degeneracy and selects the ferromagnetic state as the unique one. This is the essence of flat-band ferromagnetism, which is independent of the specific value of $U$, but depends on the filling fraction.
The illustration of polarized clusters and the corresponding wavefunction is shown in Fig.~\ref{fig:mapping_scheme}. Namely, particles with the same flavor tend to be linked together to form a cluster to minimize energy. And this is the origin of flat-band ferromagnetism~\cite{MaksymenkoPRL}. 

This energy-minimization principle allows a direct mapping from the quantum Hubbard model to a classical statistical geometric site-percolation problem~\cite{Mielke-Tasaki, MaksymenkoPRL}. 
The mapping is built upon the flat-band mechanism and is therefore independent of $U$. Under the assumptions that (i) each cell is either empty or singly occupied, and (ii) electrons in adjacent cells are in a symmetric spin state, the enumeration of all ground states of the Hubbard model on the kagome lattice maps to a percolation problem. Note that the centers of the hexagonal trapping cells of the kagome lattice align in a regular triangular lattice; each occupied trapping cell in the quantum system corresponds to an occupied site on this effective triangular lattice.  
All the ground states can be enumerated by the geometric configurations when distributing $n(\leq\mathcal{N})$ particles into $\mathcal{N}$ traps. The state composed of ferromagnetic clusters corresponds to the geometric configuration of linked trapping cells. 
The ferromagnetic transition corresponds to the emergence of such clusters.
Figure.~\ref{fig:mapping_scheme} shows the depiction of the mapping. 
 
One needs to note that the flat band of the kagome lattice is not only spanned by the ring states (Fig.~\ref{fig:mapping_scheme}), but also by additional non-contractable loop states~\cite{Bergman-PRB2008}, under periodic boundary conditions. However, non-contractable loop states do not affect the mapping to the percolation problem. The reason is that the non-contractable loop states are compatible with the ring states, where both wavefunctions have alternating signs along successive sites. However, only ring states have contributions to ferromagnetism, as the energy-minimizing principle works for the overlapping among trapping cells. Therefore, the critical particle concentration $p_{\pm}$ we found in Sec.~\ref{sec:transition} is actually one lower bound for the onset of ferromagnetism.

\begin{figure*}[htpb]
\begin{center}
\includegraphics[width=2\columnwidth]{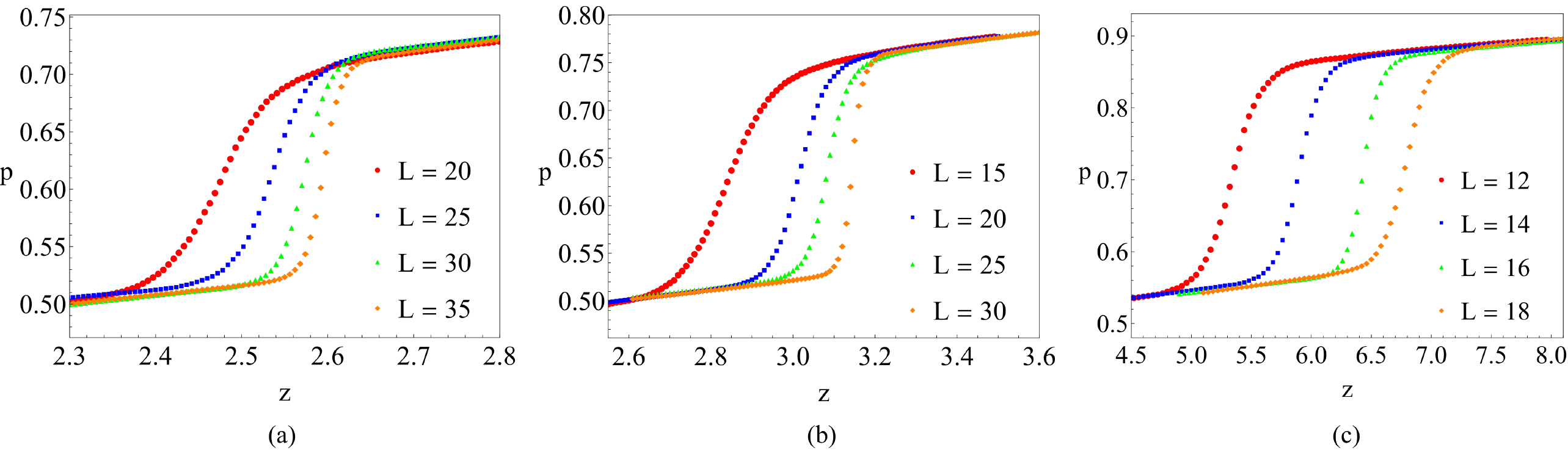}
\end{center}
\caption{Particle concentration $p$ as a function of fugacity $z$ in the grand canonical ensemble for various system sizes $|\Lambda|=L\times L$  for the Hubbard model with (a) SU$(3)$, (b) SU$(4)$, and (c) SU($10$) symmetries.}
\label{fig:p_vs_z}
\end{figure*}

This mapping was first noted by Mielke and Tasaki~\cite{Mielke-Tasaki}, and was later implemented by Monte Carlo simulations for the decorated Tasaki lattice~\cite{MaksymenkoPRL} and its SU($N$) extensions~\cite{Ruijin}. 
However, this differs from the standard percolation~\cite{Stauffer,Isichenko} due to an additional ``spin'' degeneracy for each cluster $C$ of size $|C|$. This degeneracy is given by the dimension of the fully symmetric irreducible representation of the SU($N$) group~\cite{Georgi-book,Katsura-nagaoka-SU(N)}:
\begin{equation}
d_{\text{SU}(N)}(|C|)=\frac{(N+|C|-1)!}{|C|!(N-1)!}.
\label{eq:deg}
\end{equation}
For a traditional electron system with SU($2$) symmetry, Eq.~\eqref{eq:deg} is reduced to $|C|+1$, which equals $2S+1$, where $S$ is the total spin of this cluster and $S=|C|/2$ in a magnetic cluster.  
Given the degeneracy of each cluster, each geometric configuration $q$ on the triangular lattice constituted by a set of $M_q$ distinct clusters $\{C_i, i=1,2,\cdots, M_q\}$, is assigned a statistical weight $W(q)$:
\begin{equation}
W(q)=\prod_{i=1}^{M_q}e^{\mu |C_i|}d_{\text{SU}(N)}(|C_i|),
\label{eq:weight}
\end{equation}
where the fugacity $z=e^{\mu}$ is introduced to tune the particle number in a grand-canonical ensemble. Due to the nontrivial weight originating from quantum mechanics, it is called Pauli-correlated percolation~\cite{MaksymenkoPRL}.  

Finally, we list the observables needed in the following calculation. In the context of percolation, the expectation value of an operator $O$ is a weighted average over all geometric configurations:
\begin{equation}
\langle O\rangle=\frac{\sum_q O(q) W(q)}{\sum_q W(q)}.
\label{expectation}
\end{equation}
Concerning paramagnetic-ferromagnetic transition, the most important observable is the square of the total spin  $\mathbf{S}^2$, which is the quadratic Casimir operator of the SU($N$) group. For a cluster of size $|C|$, the eigenvalue of $\mathbf{S}^2$ is given by~\cite{Ma-book}:
\begin{equation}
S^2(|C|)=\frac{(N-1)(N+|C|)|C|}{2N}.
\label{sc}
\end{equation}

\section{Para-ferro magnetic transition}
\label{sec:transition}

After mapping the quantum Hubbard model on a kagome lattice to a classical $N$-state statistical Pauli correlated site-percolation problem on an $L\times L$ triangular lattice, we perform Monte Carlo simulations, using the Metropolis algorithm with importance sampling of nontrivial weights. 
As we demonstrated in Sect.~\ref{sec:Mapping}, the mapping and subsequent simulations do not depend on the specific value of $U$; instead, they rely solely on the particle density.
The simulation is conducted on a triangular lattice with periodic boundary conditions, where the system comprises $\mathcal{N}$ trapping cells (equivalent to the flat-band degeneracy). We consider $n \le \mathcal{N}$ particles distributed among these cells, with each cell restricted to being either empty or singly occupied. In each Monte Carlo step, a proposed new configuration $q'$ is accepted with Metropolis probability $\min[1, W(q')/W(q)]$, where $q$ denotes the current configuration and $W(q)$ is the nontrivial statistical weight defined in Eq.~\eqref{eq:weight}. Hoshen-Kopelman algorithm~\cite{Hoshen-Kopelman} is employed in cluster labeling.

To determine the ground state phase diagram, we employ a two-step numerical strategy.
First, we perform grand-canonical ensemble simulations using the replica exchange Monte Carlo method~\cite{exchangeMC} to improve sampling efficiency and accuracy. In this setup, a trial move consists of randomly selecting a site and flipping its occupancy (i.e., removing a particle if the site is occupied, or adding one if it is empty). These simulations yield the relation between the particle concentration $p \equiv n/\mathcal{N}$ and the fugacity $z = e^{\mu}$ (or chemical potential $\mu$).

As shown in Fig.~\ref{fig:p_vs_z}, the curve exhibits a discontinuous rapid rise from $p_-$ to $p_+$ at a critical fugacity $z_c$, which sharpens with increasing system size. The rapid rise allows for the estimation of the critical concentration $p(z_c)$ bounded by $p _-$ and $p_+$. The values of densities $p_-$ and $p_+$ demarcate distinct thermodynamic phases. For $p < p_-$, the probability of forming large same-color clusters vanishes in the thermodynamic limit, corresponding to a paramagnetic phase in the original Hubbard model. Conversely, for $p > p_+$, clusters of the same color can connect and reach a macroscopic size, signaling a ferromagnetic phase of the original Hubbard model. The region between $p_-$ and $p_+$ thus represents a regime of phase coexistence.

\begin{figure}[htpb]
\begin{center}
\includegraphics[width=0.7\columnwidth]{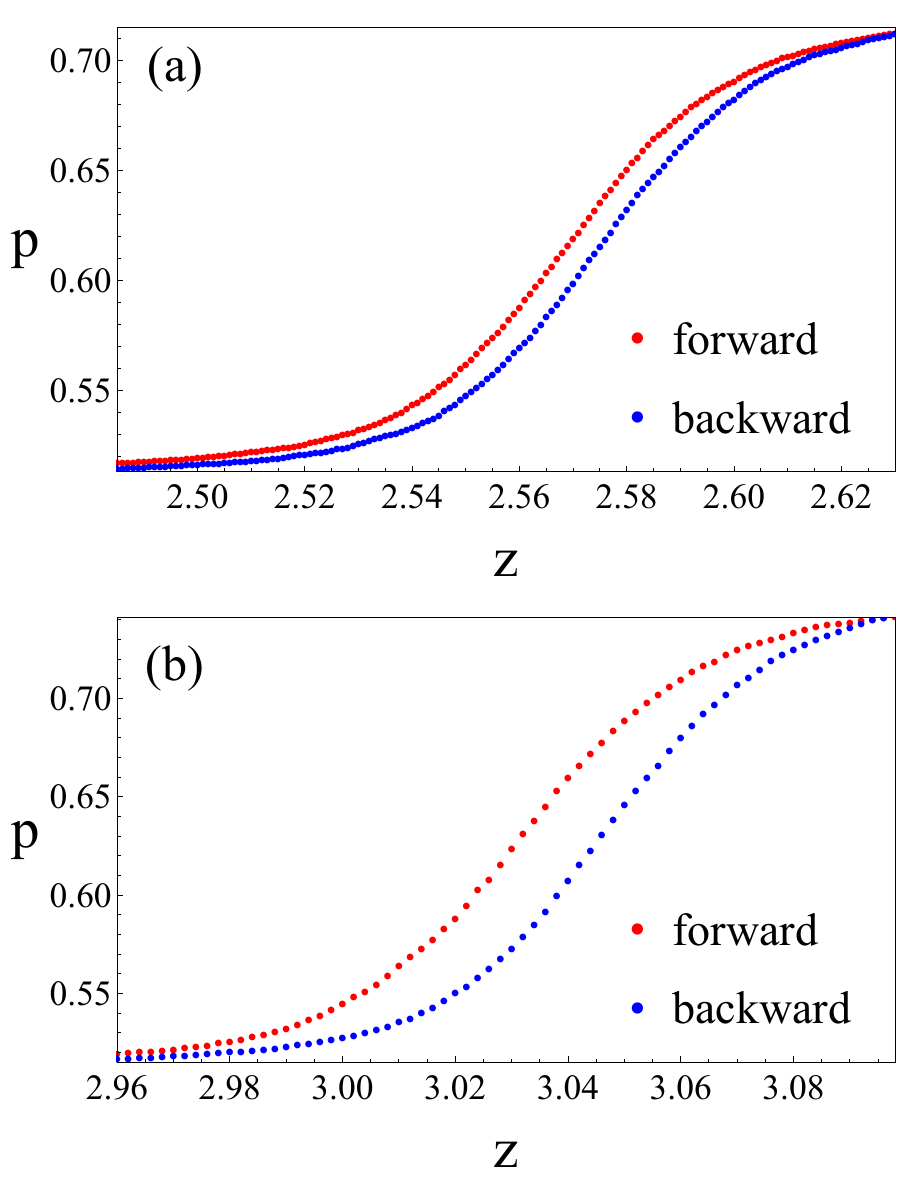}
\end{center}
\caption{Hysteresis loops obtained from exchange Monte Carlo simulations in the grand canonical ensemble by sweeping $z$ forward and backward for (a) SU(3) and (b) SU(4) systems, where lattice extension $L=30$.}
\label{fig:loop}
\end{figure}

\begin{figure*}[htpb]
\begin{center}
\includegraphics[width=2\columnwidth]{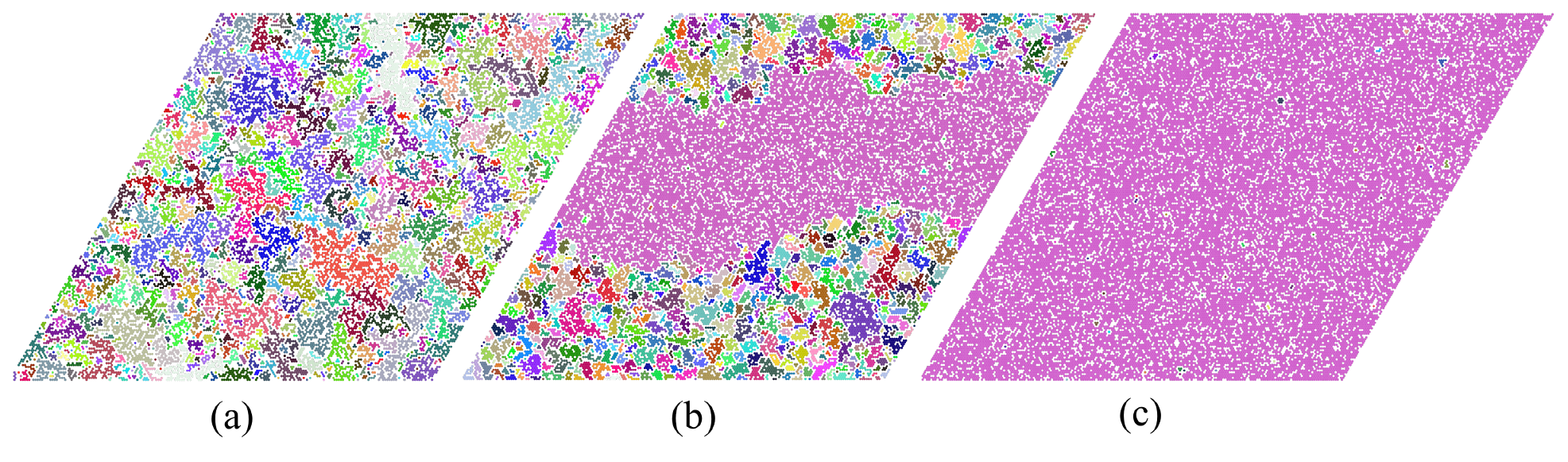}
\end{center}
\caption{Snapshots of typical configurations for Pauli-correlated percolation for small deviations from critical concentration, by Monte Carlo simulation for SU($3$) Hubbard model. Snapshots for concentrations at (a) $p_1=0.48$ (paramagnetic), (b) $p_2=0.62$ (phase-separated), and (c) $p_3=0.73$ (ferromagnetic). Lattice extension $L=200$. Empty sites are white, and the largest cluster is magenta.}
\label{fig:snapshots}
\end{figure*}

We further examine the nature of the paramagnetic-ferromagnetic transition. When sweeping the $z$ values, we use the last configuration for a $z$ value in the Markov chain as the initial configuration for simulations at the next larger $z$. We further sweep $z$ in the opposite direction and find a hysteresis loop (Fig.~\ref{fig:loop}), a telltale sign of first-order transition. The first-order nature of this transition is also corroborated by direct visual inspection of the spatial configurations obtained from canonical ensemble simulations, as shown in Fig.~\ref{fig:snapshots}, where a phase separation region exists.

\begin{figure*}[htbp]
\begin{center}
\includegraphics[width=2\columnwidth]{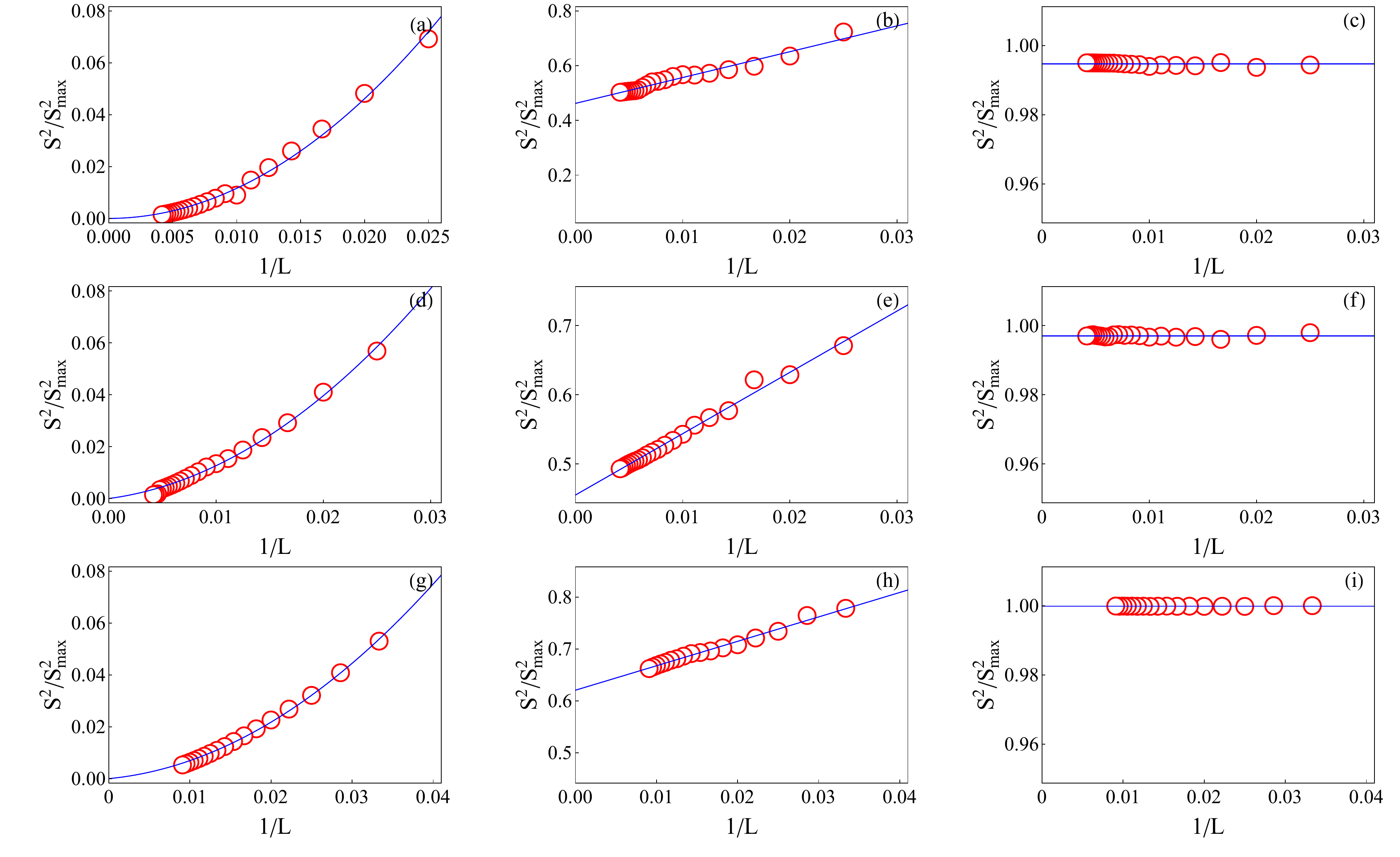}
\end{center}
\caption{Finite-size scaling of the normalized macroscopic magnetic moment, $\langle S^2 \rangle / S^2_{\max}$, for the Hubbard model on a two-dimensional kagome lattice. Results are shown for SU($3$) [(a)-(c)], SU($4$) [(d)-(f)], and SU($10$) [(g)-(i)] symmetries, with maximum system sizes of $L=240$, $200$, and $110$, respectively. The corresponding particle concentrations $p$ are: (a) $0.50$, (b) $0.62$, (c) $0.72$ for SU($3$); (d) $0.51$, (e) $0.65$, (f) $0.76$ for SU($4$); and (g) $0.52$, (h) $0.76$, (i) $0.87$ for SU($10$).}
\label{fig:results_scaling}
\end{figure*}

Second, to locate the paramagnetic-ferromagnetic transition, we perform canonical ensemble Monte Carlo simulations at fixed particle concentrations $p$. In this ensemble, particle number conservation is enforced by generating trial moves through the exchange of occupancy between two randomly selected sites. A proposed new configuration is accepted according to the standard Metropolis criterion.
We compute the macroscopic magnetic moment $\langle S^2 \rangle$ for various system sizes. Its finite-size scaling behavior, shown in Fig.~\ref{fig:results_scaling}, clearly distinguishes the two phases. In the thermodynamic limit, the normalized moment $\langle S^2 \rangle / S^2_{\max}$ extrapolates to zero for $p < p_-$ (paramagnetic phase), while it converges to a finite value for $p > p_+$ (ferromagnetic phase). Here, $S^2_{\max}$ is the maximum possible moment, corresponding to a state where all particles align within a single percolating cluster.

As shown in Fig.~\ref{fig:results_scaling} (a, d, g), the normalized macroscopic magnetic moment vanishes in the thermodynamic limit at particle densities $p = 0.50$, $0.51$, and $0.52$ for the SU($3$), SU($4$), and SU($10$) models, respectively. This scaling behavior, corroborated by the corresponding configuration snapshots in Fig.~\ref{fig:snapshots} (a), confirms a paramagnetic ground state at these low densities.

Upon increasing the density to $p = 0.62$ (SU($3$)), $0.65$ (SU($4$)), and $0.76$ (SU($10$)) [Fig.~\ref{fig:results_scaling} (b, e, h)], the normalized moment extrapolates to a finite value in the thermodynamic limit, signaling the onset of ferromagnetism. In this regime, the system enters a phase-separated phase, where macroscopic ferromagnetic domains coexist with a paramagnetic background, as visualized in the snapshot for SU($3$) case in Fig.~\ref{fig:snapshots} (b). 
This phase-separation behavior, absent in standard site percolation (where the statistical weight $W(q)=1$), arises in our correlated percolation problem because the nontrivial weight $W(q) \equiv \exp[\ln W(q)]$ introduces an effective entropic interaction ($\ln W(q)$)~\cite{MaksymenkoPRL}. Such a mechanism, previously identified on the square lattice~\cite{MaksymenkoPRL}, is now further observed here on the triangular lattice for SU($N$) systems, and also in SU($2$) system~\cite{YiLi-PRB2021}.

When the density is further enhanced to $p = 0.72$ (SU($3$)), $p = 0.76$ (SU($4$)), and $p = 0.87$ (SU($10$))[Fig.~\ref{fig:results_scaling} (c, f, i)], the normalized moment saturates to a value close to unity ($\langle S^2 \rangle / S^2_{\max} \approx 1$) and becomes independent of system size. This indicates the establishment of a fully developed ferromagnetic phase that percolates throughout the entire system, as depicted in Fig.~\ref{fig:snapshots} (c).

\begin{figure}[htpb]
\begin{center}
\includegraphics[width=0.8\columnwidth]{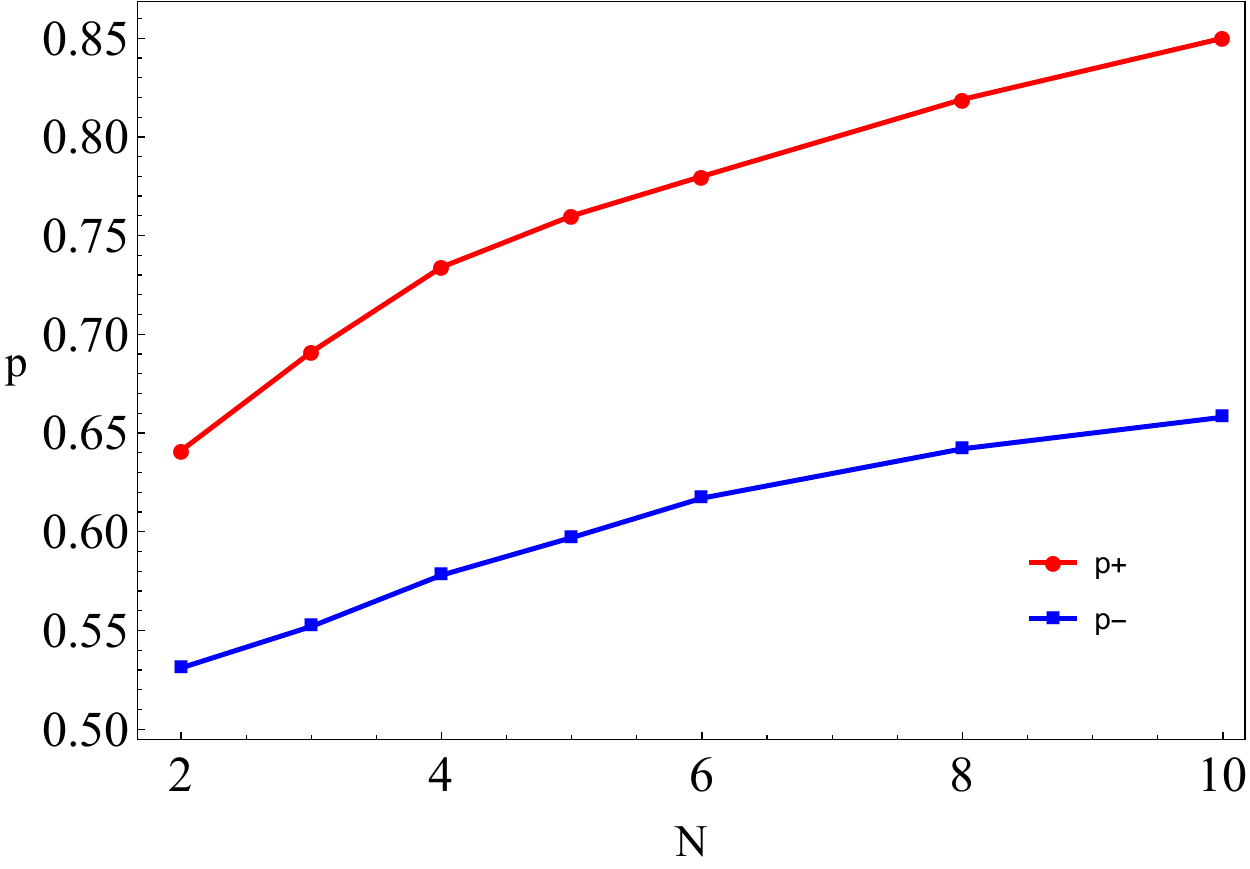}
\end{center}
\caption{Phase diagram in the $p$-$N$ space. The paramagnetic phase is found below $p_-$; the ferromagnetic phase, above $p_+$; and the phase-separated phase, between $p_-$ and $p_+$.}
\label{fig:p_N}
\end{figure}

Based on the numerical results, we construct the ground-state phase diagram of the SU($N$) Hubbard model on the kagome lattice as shown in Fig.~\ref{fig:p_N}. The paramagnetic-ferromagnetic transition is first order, characterized by a discontinuous rapid rise between two critical densities, $p_-$ and $p_+$. The phase diagram comprises three distinct regimes: (i) a paramagnetic phase lies below $p_-$; (ii) an unsaturated ferromagnetic phase lies above $p_+$, which percolates throughout the system; and (iii) a phase-separated regime occupies the region between $p_-$ and $p_+$, where ferromagnetic domains coexist with a paramagnetic background. Notably, both densities $p_-$ and $p_+$ increase with the flavor number $N$ and are larger than the critical filling $p_c=0.5$ for standard site percolation on the triangular lattice~\cite{Grimmett-percolation}, as a direct consequence of the stronger effective entropic repulsion in systems with higher 
SU($N$) symmetry.

\section{Conclusion}
\label{sec:Conclusion}

We have studied the paramagnetic-ferromagnetic transition in the repulsive SU($N$) Fermi Hubbard model on the geometrically frustrated kagome lattice for $N = 3, 4, 10$. With the concentration restriction $p \equiv n / \mathcal{N} \le 1$, where $n$ is the number of particles and $\mathcal{N}$ the number of trapping cells (also equal to the degeneracy of the single-particle flat band), the ground states of the quantum model can be exactly mapped to the geometric configurations of a classical $N$-state Pauli-correlated site-percolation model on a triangular lattice. This mapping enables us to circumvent the fermion sign problem and perform large-scale, efficient classical Monte Carlo simulations.

Our numerical analysis reveals a first-order paramagnetic-ferromagnetic phase transition, characterized by a discontinuous rapid rise between two critical densities, $p_-$ and $p_+$. Via finite-size scaling, we have estimated the values of $p_-$ and $p_+$: $p_- = 0.55(2)$ and $p_+ = 0.69(1)$ for SU($3$); $p_- = 0.57(8)$ and $p_+ = 0.73(4)$ for SU($4$); $p_- = 0.59(7)$ and $p_+ = 0.76(0)$ for SU($5$); $p_- = 0.61(7)$ and $p_+ = 0.78(0)$ for SU($6$); $p_- = 0.64(3)$ and $p_+ = 0.81(9)$ for SU($8$); $p_- = 0.65(8)$ and $p_+ = 0.85(0)$ for SU($10$).

These results conclusively show the increase of both $p_-$ and $p_+$ with the flavor number $N$, demonstrating that achieving ferromagnetism requires higher particle densities in systems with larger SU($N$) symmetry. This trend is a direct manifestation of the stronger effective entropic repulsion arising from the increased state degeneracy, which suppresses the growth of macroscopic ferromagnetic clusters and thereby elevates the critical threshold for long-range order.

In summary, our work provides quantitative insights into the rich interplay between lattice geometry, high internal symmetry, and many-body interactions. It establishes the mapped percolation framework as a powerful tool for studying sign-problem-free SU($N$) physics and further solidifies the kagome lattice as a paradigmatic platform for discovering exotic correlation-driven phenomena.

\section{Acknowledgement}
We would like to acknowledge earlier extensive collaborations on related subjects with Rui-Jin Liu and Wei Zhang. These collaborations laid the foundation for the present work. W. N. was supported by the National Natural Science Foundation of China (NSFC) under Grant No. 12174273.


%

\end{document}